\documentclass[aip,reprint]{revtex4-1}

\usepackage[
    separate-uncertainty = true]{siunitx}
\usepackage[format=plain,justification=Justified, textfont=small]{caption} %,justification=RaggedRight
\usepackage{subcaption}
\usepackage{amsmath,amssymb}
\usepackage{graphicx}% Include figure files
\usepackage{color, soul} % Highlight text

\draft % marks overfull lines with a black rule on the right

\begin{document}

% Use the \preprint command to place your local institutional report number 
% on the title page in preprint mode.
% Multiple \preprint commands are allowed.
%\preprint{}

\title{Trap-Integrated Superconducting Nanowire Single-Photon Detectors with Improved RF Tolerance for Trapped-Ion Qubit State Readout} %Title of paper

% repeat the \author .. \affiliation  etc. as needed
% \email, \thanks, \homepage, \altaffiliation all apply to the current author.
% Explanatory text should go in the []'s, 
% actual e-mail address or url should go in the {}'s for \email and \homepage.
% Please use the appropriate macro for the type of information

% \affiliation command applies to all authors since the last \affiliation command. 
% The \affiliation command should follow the other information.

\author{Benedikt Hampel}
\email[]{benedikt.hampel@nist.gov}
%\homepage[]{Your web page}
%\thanks{}
%\altaffiliation{}
\affiliation{National Institute of Standards and Technology, Boulder, Colorado 80305, USA}
\affiliation{Department of Physics, University of Colorado, Boulder, Colorado 80309, USA}

\author{Daniel H. Slichter}
\author{Dietrich Leibfried}
\author{Richard P. Mirin}
\author{Sae Woo Nam}
\author{Varun B. Verma}
\affiliation{National Institute of Standards and Technology, Boulder, Colorado 80305, USA}

% Collaboration name, if desired (requires use of superscriptaddress option in \documentclass). 
% \noaffiliation is required (may also be used with the \author command).
%\collaboration{}
%\noaffiliation

\date{\today}

\begin{abstract}
  State readout of trapped-ion qubits with trap-integrated detectors can address important challenges for scalable quantum computing, but the strong rf electric fields used for trapping can impact detector performance. Here, we report on NbTiN superconducting nanowire single-photon detectors (SNSPDs) employing grounded aluminum mirrors as electrical shielding that are integrated into linear surface-electrode rf ion traps. The shielded SNSPDs can be successfully operated at applied rf trapping potentials of up to \SI{54}{\volt\of{peak}} at \SI{70}{\mega\hertz} and temperatures of up to \SI{6}{\kelvin}, with a maximum system detection efficiency of \SI{68}{\percent}. This performance should be sufficient to enable parallel high-fidelity state readout of a wide range of trapped ion species in typical cryogenic apparatus. 
\end{abstract}

% insert suggested PACS numbers in braces on next line

\maketitle %\maketitle must follow title, authors, abstract and \pacs

% Body of paper goes here. Use proper sectioning commands. 
% References should be done using the \cite, \ref, and \label commands
%\label{}
%\subsection{}
%\subsubsection{}

The state readout of trapped-ion qubits is usually accomplished by driving an optical cycling transition with one or more laser beams and observing qubit-state-dependent fluorescence from the ion~\cite{dehmelt_1982}. Counting just a few percent of the total fluorescence photons from the ion is generally sufficient for state readout with fidelity \num{>0.999}~\cite{Wineland1998}. The ion fluorescence is at ultraviolet (UV) or blue wavelengths for nearly all species used experimentally. Most experimental setups use high-numerical-aperture fluorescence collection optics combined with photon detectors such as cameras or photomultiplier tubes to detect the presence or absence of fluorescence and thus infer the qubit state. The simultaneous requirements on field of view, numerical aperture, off-axis aberration, magnification, and physical dimensions of the collection optics present challenges when scaling this architecture to large numbers of ions. Applications such as scaled trapped-ion quantum computing~\cite{Kielpinski2002, Monroe2013, Bruzewicz.2019} may benefit from the integration of multiple photon detectors for fluorescence detection into the ion trap itself, without requiring optical elements for imaging~\cite{Slichter.2017, Todaro.2021, setzer.2021, reens.2022}. 

Superconducting nanowire single-photon detectors (SNSPDs) are promising candidates for this application, as they have very low dark count rates~\cite{wollman.2017}, high detection efficiencies~\cite{Reddy.2020}, high maximum count rates over a wide range of wavelengths, including ultraviolet wavelengths~\cite{Todaro.2021, crain.2019, Slichter.2017}, and can be fabricated as devices with many separate pixels~\cite{Wollman2019}, as desirable for parallel detection of ions in a large array of traps~\cite{Wineland1998}. Many ion trap experiments already operate at cryogenic temperatures between roughly \SI{4}{\kelvin} and \SI{10}{\kelvin}, such that the cryogenic temperatures required for SNSPD operation might not present a major obstacle. However, the strong rf electric fields used to trap ions can induce undesired rf currents in trap-integrated SNSPDs and degrade their performance~\cite{Slichter.2017, Todaro.2021}. Other approaches for trap-integrated photon detectors include single-photon avalanche photodiodes (SPADs), which can be operated in room temperature ion traps. Trap-integrated SPADs also suffer from the same rf pickup effects, however, and have much higher dark count rates and lower quantum efficiencies than SNSPDs. SPAD output pulses also cause more heating of the ion motion than SNSPD pulses due to their larger amplitude~\cite{setzer.2021, reens.2022, Todaro.2021}.

In this work, we demonstrate an improved design of a linear ion trap with an integrated SNSPD, optimized for detection of fluorescence from Ca$^+$ ions at \SI{397}{\nano\meter} and shielded from the trapping rf fields by an aluminum mirror under the device that also contributes to improved detection efficiency. Based on the measured system performance and the results in Ref.~\onlinecite{Todaro.2021}, these devices should enable high-fidelity qubit state readout of a trapped Ca$^+$ ion under realistic trap operating conditions (temperature and trapping rf).

The trap electrode design is identical to that in Ref.~\onlinecite{Todaro.2021}, but the SNSPDs differ in two major ways. First, we use niobium titanium nitride (NbTiN) instead of molybdenum silicide (MoSi) as the superconducting material for the SNSPDs~\cite{Dorenbos.2008,Tanner.2010,Miki.2013,Chang.2021}, thus increasing the SNSPD critical currents by roughly a factor of \num{10} and enabling operation at higher temperatures of up to \SI{7}{\kelvin}~\cite{Gourgues.2019}. These features reduce the sensitivity of the SNSPD performance to the applied trap rf potentials. Second, a grounded aluminum mirror fabricated below the SNSPD provides electrical shielding and increases the detection efficiency of the SNSPD. We characterize the performance of SNSPDs in the presence of realistic rf trapping potentials suitable for trapping a wide range of ion species and determine their system detection efficiency and dark count performance.

\begin{figure}%[b]
    \includegraphics[width=0.45\textwidth]{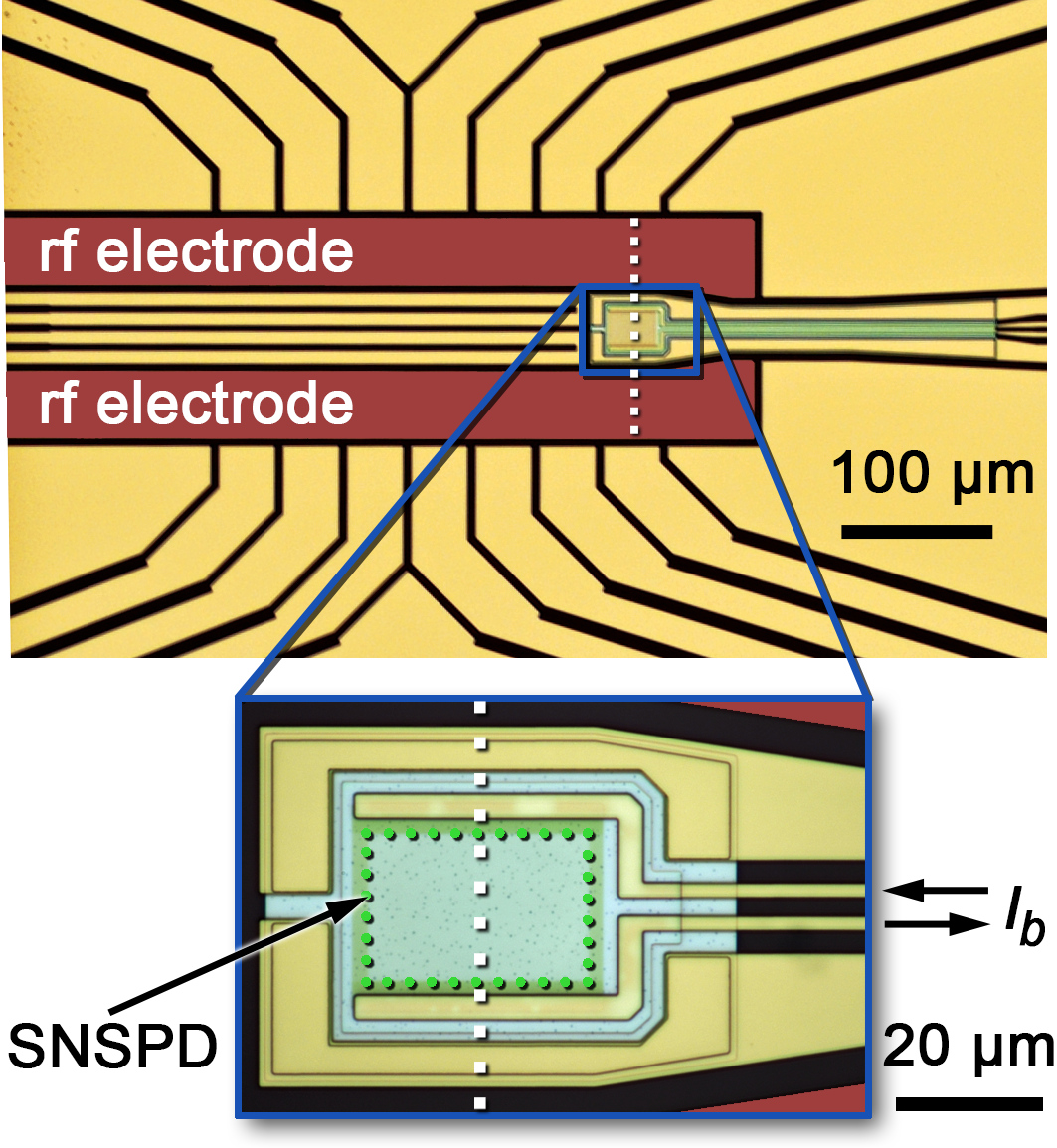}
    \caption{Photomicrograph of the linear ion trap with integrated SNSPD. The rf trapping potentials are applied to the labeled and false-colored  rf electrodes. All other non-SNSPD electrodes are used for dc trapping potentials. The magnified inset shows the SNSPD on top of a small aluminum mirror, while a larger aluminum mirror spanning the region beneath the leads of the SNSPD can be seen in the larger picture. Both mirrors appear light blue in the gaps between gold electrodes, while the black gaps show the substrate. The green dotted line borders the active area of the SNSPD. The white dashed lines indicate the plane of the 2D simulations in Fig.~\ref{COMSOL}. The bias current $I_b$ is applied to the leads of the SNSPD.}
    \label{iontrap}
\end{figure}

An ion trap with integrated SNSPD is depicted in the photomicrograph in Fig.~\ref{iontrap}. We designed the shielding mirror by performing two-dimensional finite-element electromagnetic simulations of the trap geometry in the plane normal to the trap surface along the vertical white dotted line in Fig.~\ref{iontrap}. All dielectric and metal layers on the wafer, including the gold trap electrodes, were included in the simulation. We analyzed the electric field at the SNSPD when an rf potential was applied to the rf electrodes of the ion trap (false colored red in Fig.~\ref{iontrap}). The simulated rf frequency of \SI{70}{\mega\hertz} is suitable for trapping a wide range of ions including the target Ca$^+$ species.

\begin{figure}[b]
    \includegraphics[width=0.49\textwidth]{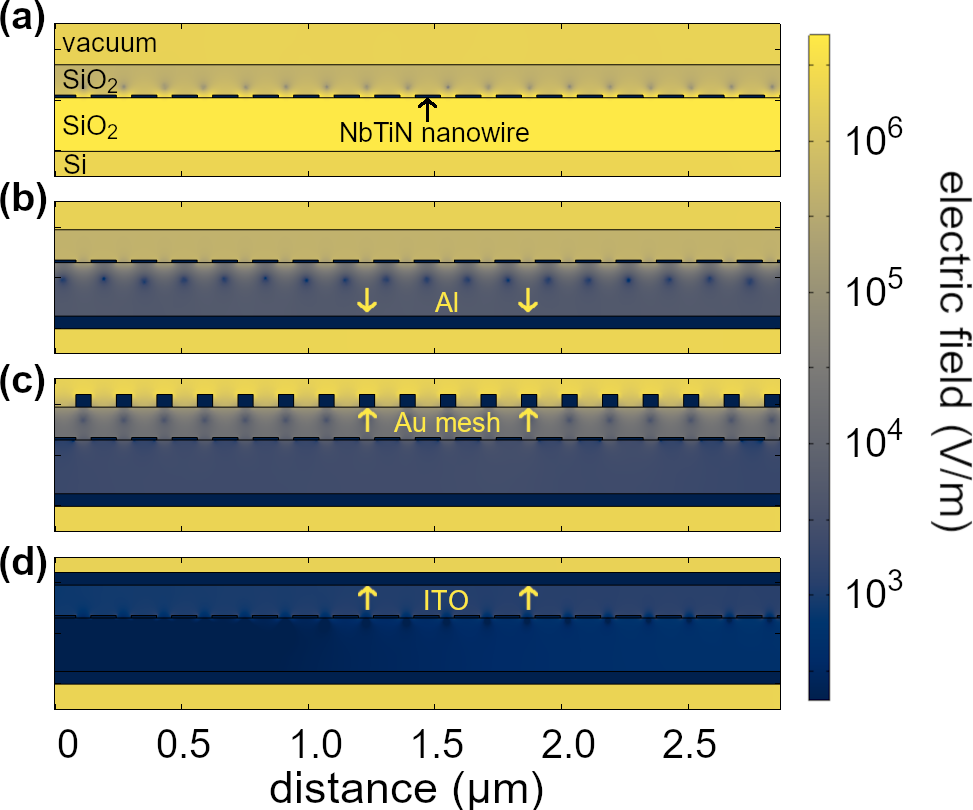}
    \caption{Numerically simulated electric field amplitude from an rf potential of \SI{100}{\volt\of{peak}} at \SI{70}{\mega\hertz} applied to the rf electrodes of the ion trap, viewed for a small cross-section of the SNSPD and for different shielding configurations. (a) SNSPD nanowire embedded in SiO\textsubscript{2} without shielding. (b) Grounded aluminum mirror under the nanowire. (c) Grounded gold mesh on top of the stack in addition to the grounded aluminum mirror. (d) Grounded ITO layer instead of the gold mesh. }
    \label{COMSOL}
\end{figure}

The results of the simulations are depicted in Fig.~\ref{COMSOL}. The plot color represents the rf electric field amplitude on a logarithmic scale. Larger rf electric fields at the position of the individual parts of the nanowire cause larger induced rf currents and correspondingly larger impact on the SNSPD performance~\cite{Slichter.2017, Todaro.2021}. Fig.~\ref{COMSOL}(a) simulates an SNSPD without shielding, similar to the device described in Ref.~\onlinecite{Todaro.2021}. The illustrated layer stack from bottom to top consists of the silicon wafer, a \SI{212}{\nano\meter}-thick SiO\textsubscript{2} layer, \SI{10}{\nano\meter}-thick and \SI{100}{\nano\meter}-wide NbTiN nanowire on a pitch of \SI{160}{\nano\meter}, a \SI{130}{\nano\meter}-thick SiO\textsubscript{2} layer, and vacuum above. In Fig.~\ref{COMSOL}(b), a grounded aluminum mirror is added between the silicon wafer and the lower SiO\textsubscript{2} layer to shield the SNSPD. The rf electric field around the nanowire cross-sections is reduced by roughly an order of magnitude relative to the unshielded case, suggesting greater immunity to the trapping rf fields. Additional grounded conductors on top, either a \SI{50}{\nano\meter}-thick gold mesh with width and pitch equal to the SNSPD gaps (Fig.~\ref{COMSOL}(c)) or a \SI{50}{\nano\meter}-thick indium tin oxide (ITO) layer (Fig.~\ref{COMSOL}(d)), further improve the shielding of the SNSPD against rf signals, while still transmitting most of the ion fluorescence photons to the nanowire. The top electrical shielding layer both protects the SNSPD from trapping rf as well as shielding the ion from electric fields due to SNSPD pulses. 

The mirror also serves as part of an optical stack in combination with the dielectric layers. Based on rigorous coupled-wave analysis (RCWA)~\cite{Moharam.1995}, the mirror increases the absorption of normally incident photons in the nanowire by a factor of two when the SiO\textsubscript{2} thicknesses are chosen appropriately; aluminum was chosen for its higher UV/blue reflectivity than gold.

Based on the simulation results, we fabricated traps with integrated SNSPDs on top of grounded aluminum mirrors but without top shielding layers for simplicity, as shown in Fig.~\ref{COMSOL}(b). The SNSPD geometry and dielectric stack were optimized for absorption at the Ca\textsuperscript{+} ion fluorescence wavelength of \SI{397}{\nano\meter}. All devices were fabricated in the NIST Boulder Microfabrication Facility. 

The devices were fabricated on high-resistivity (\SI{>20}{\kilo\ohm\cm}) silicon wafers on which \SI{150}{\nano\meter} of SiO\textsubscript{2} were deposited in a plasma-enhanced chemical vapor deposition (PECVD) process. A \SI{50}{\nano\meter} electron-beam evaporated aluminium layer to form the mirrors was patterned in a lift-off process and covered with \SI{212}{\nano\meter} of SiO\textsubscript{2}. This was followed by a \SI{10}{\nano\meter}-thick NbTiN film deposited by co-sputtering at \SI{500}{\celsius} on the lower SiO\textsubscript{2} layer. Nanowire meanders were patterned in the NbTiN by optical and electron-beam lithography combined with reactive-ion etching with sulfur hexafluoride (SF\textsubscript{6}). The SNSPD active area measures \SI{20 x 30}{\micro\meter}. The nanowire width was determined with a scanning electron microscope as \SI{90}{\nano\meter} with a pitch of \SI{180}{\nano\meter}. Vias were then etched through the SiO\textsubscript{2} bottom layer in a reactive-ion etching process with trifluoromethane (CHF\textsubscript{3}), followed by a \SI{20}{\nano\meter}-thick gold layer, to connect to the buried aluminum mirror electrically. 

The rf (red) and dc (gold) electrodes of the ion trap in Fig.~\ref{iontrap} provide the potentials for trapping ions when driven appropriately. They were patterned, together with the SNSPD leads, in a lift-off process from a \SI{350}{\nano\meter} evaporated gold layer, making electrical contact with the aluminum mirror layer in regions where vias were previously opened.  Finally, a \SI{130}{\nano\meter}-thick SiO\textsubscript{2} cap was deposited on top of the SNSPD to protect the nanowire during further processing and limit degradation due to oxidation. This layer was deposited across the entire wafer in a PECVD process and was patterned to cover only the SNSPD using a CHF\textsubscript{3} reactive-ion etching process.

We fabricated and tested two different mirror sizes (large and small), with the mirrors either electrically floating or grounded via the trap electrodes, to compare shielding performance. The small mirror can be seen in the magnified inset of Fig.~\ref{iontrap} and covers the area under the SNSPD and a small region around it. The larger mirror is depicted in the top photomicrograph of Fig.~\ref{iontrap}. It was fabricated to cover a larger area, including under the electrical leads of the SNSPD for several hundred \SI{}{\micro\meter} away from the rf electrode.

\begin{figure*}[t]
  \centering
  \begin{subfigure}[b]{0.47\textwidth}
    \centering
    \includegraphics[width=1.0\textwidth]{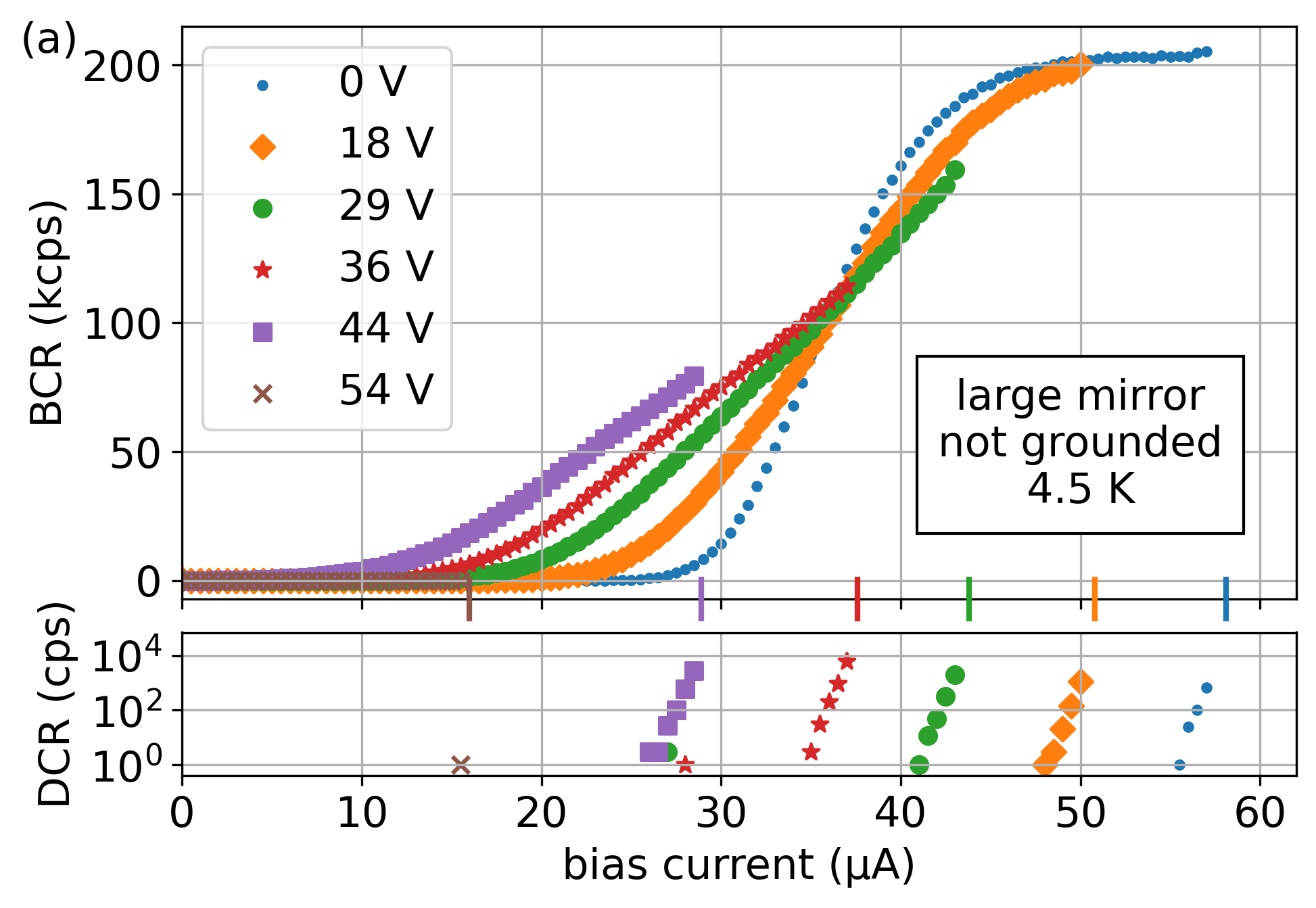}
    %\caption{Large mirror, grounded, T = \SI{4.5}{\kelvin}}
    \label{measurement:LgGnd4p5}
    \vspace{-5mm}
  \end{subfigure}  
  \begin{subfigure}[b]{0.47\textwidth}
    \centering
    \includegraphics[width=1.0\textwidth]{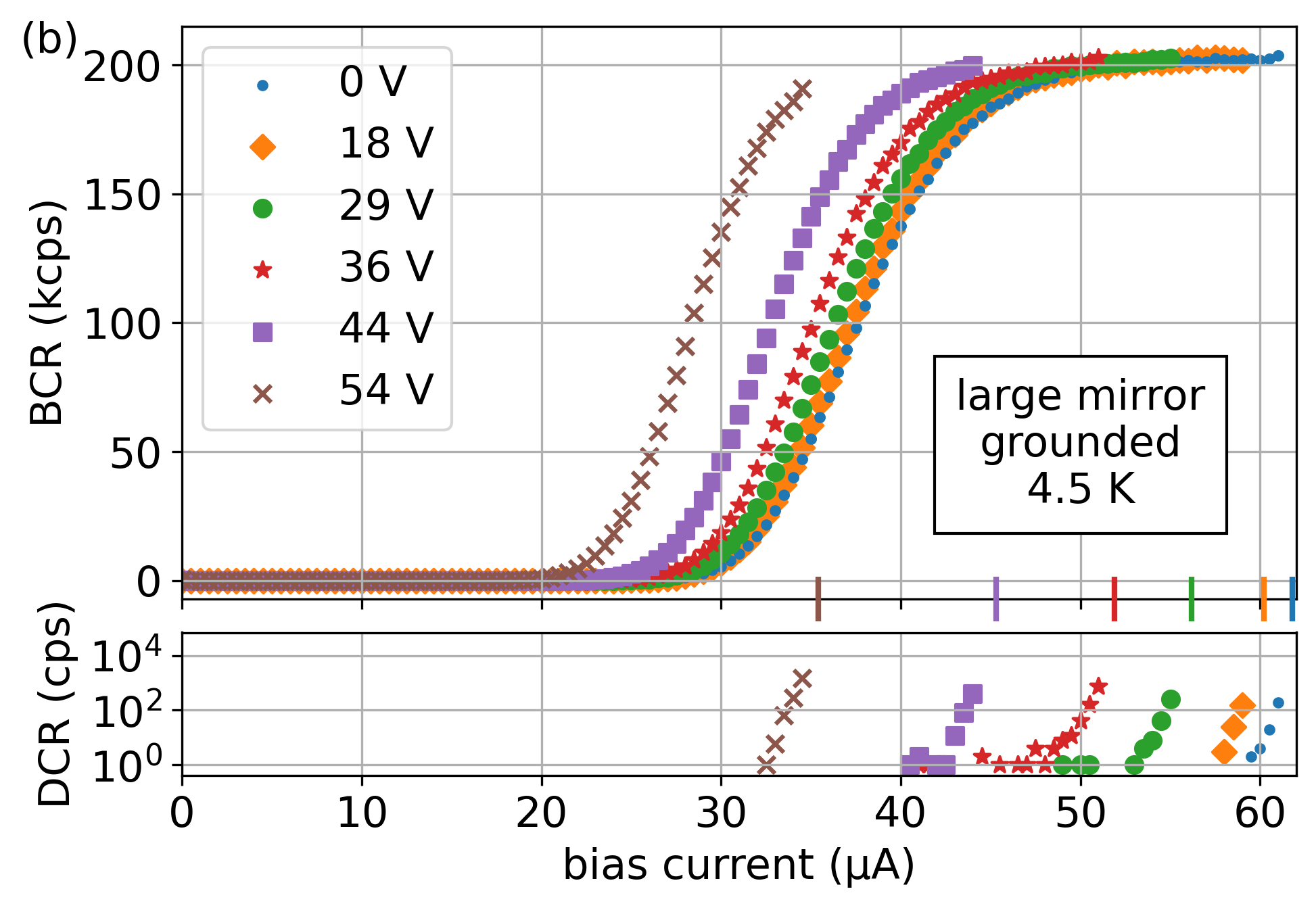}
    %\caption{Large mirror, grounded, T = \SI{6}{\kelvin}}
    \label{measurement:LgGnd6}
    \vspace{-5mm}
  \end{subfigure}
  \begin{subfigure}[b]{0.47\textwidth}
    \centering
    \includegraphics[width=1.0\textwidth]{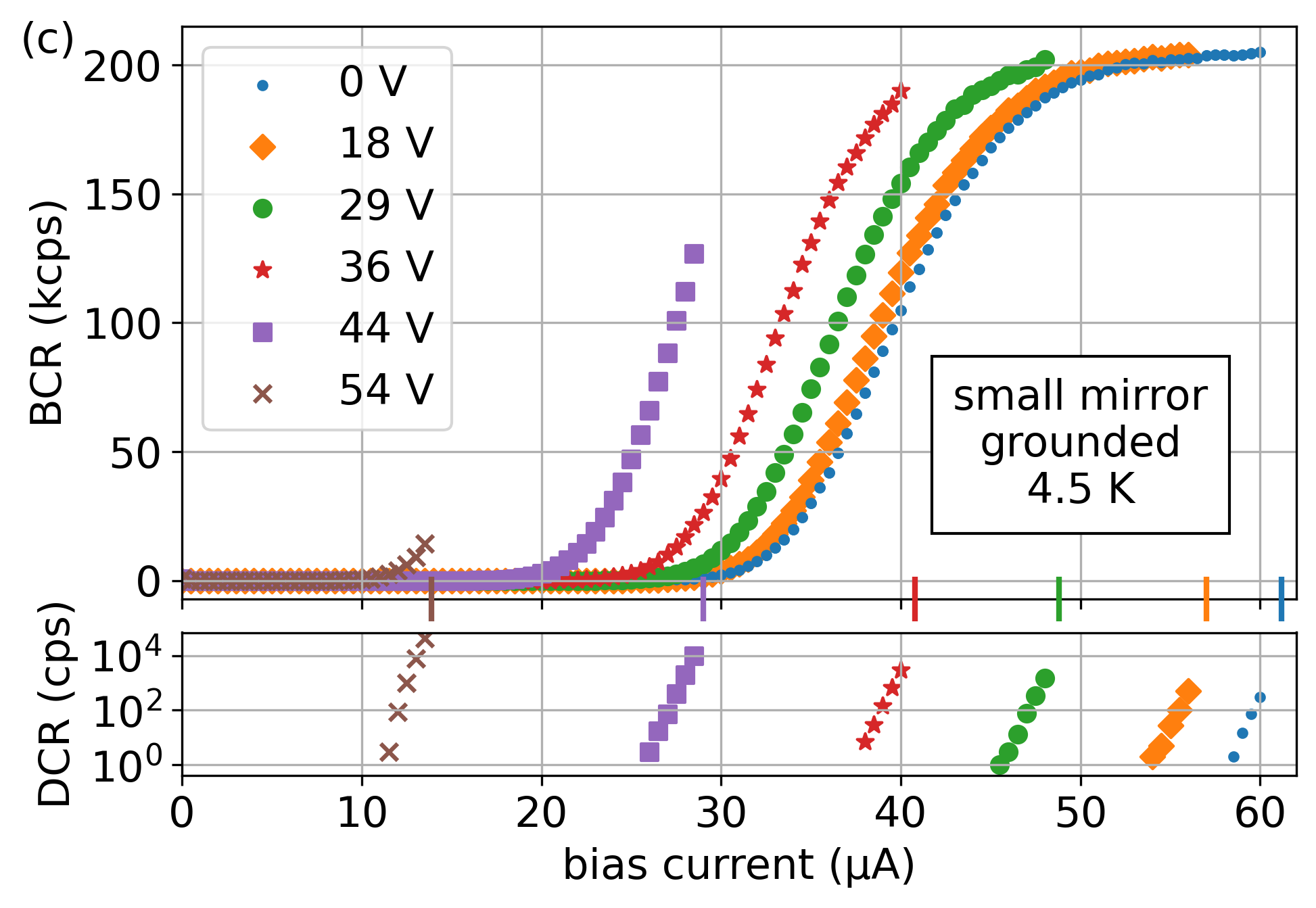}
    %\caption{Large mirror, not grounded, T = \SI{4.5}{\kelvin}}
    \label{measurement:Lg4p5}
    \vspace{-5mm}
  \end{subfigure}  
  \begin{subfigure}[b]{0.47\textwidth}
    \centering
    \includegraphics[width=1.0\textwidth]{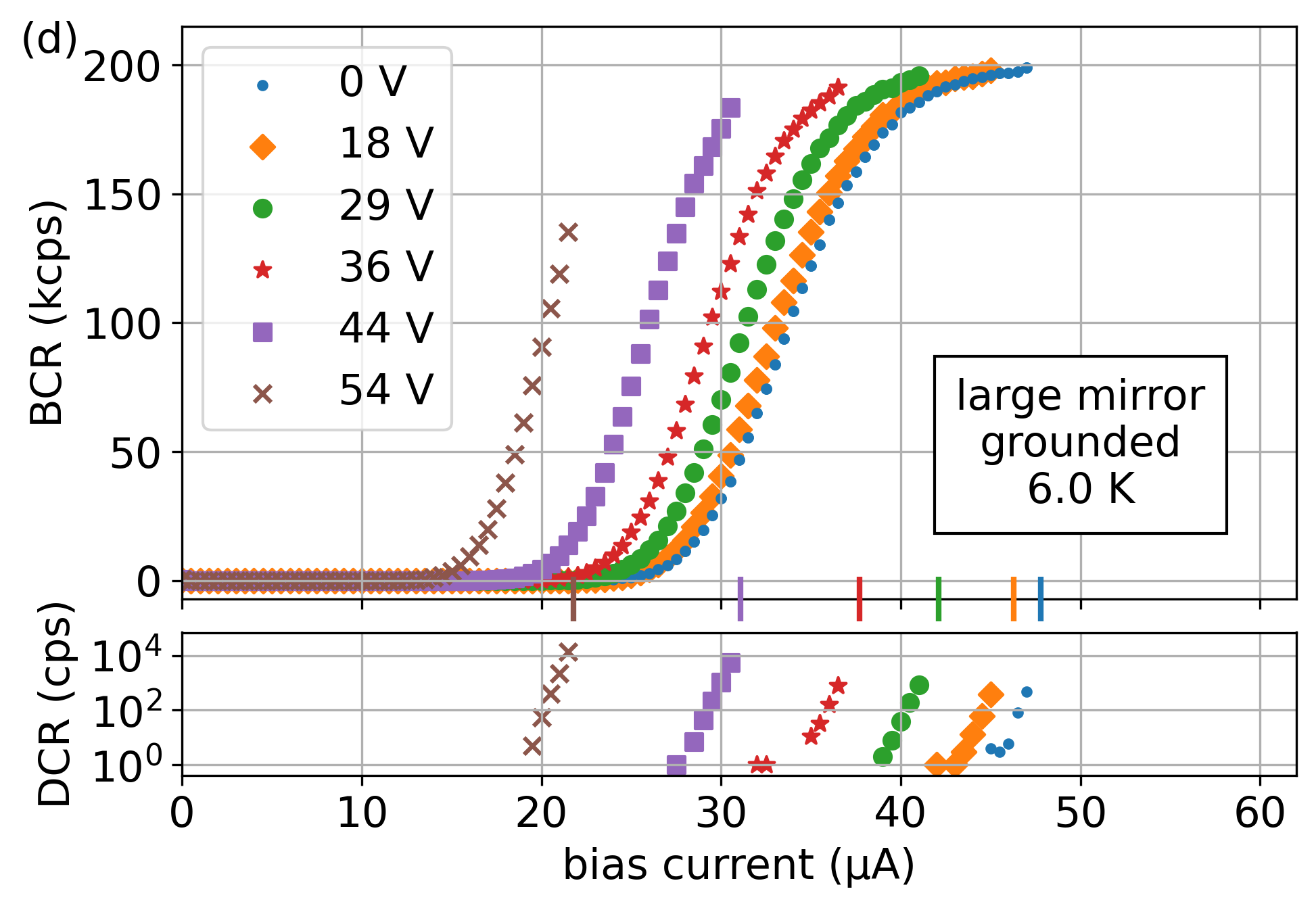}
    %\caption{Small mirror, grounded, T = \SI{4.5}{\kelvin}}
    \label{measurement:SmGnd4p5}
    \vspace{-5mm}
  \end{subfigure}
  \caption{Bright count rate (BCR) and dark count rate (DCR) in counts per second (cps) as a function of bias current for different rf voltage peak amplitudes on the ion trap at a temperature of \SI{4.5}{\kelvin} for devices with a large mirror that is (a) not connected to ground and (b) connected to ground. Comparison to device with (c) a smaller grounded mirror at \SI{4.5}{\kelvin} and (d) the same device as in (b) operated at \SI{6}{\kelvin}. The switching currents for each value of applied rf are indicated with correspondingly colored vertical markers on the horizontal axes. BCR values shown are corrected for DCR. DCR is plotted on logarithmic vertical axes.}
  \label{measurement}
\end{figure*}

The measurements of the performance and the rf tolerance of the trap-integrated SNSPDs were performed in a variable-temperature cryostat with a base temperature of \SI{2.6}{\kelvin}. Cryogenically-operated LEDs with a peak wavelength of \SI{395}{\nano\meter} provided flood illumination of the samples. The rf electrodes of the ion traps were connected to rf resonators with a voltage step-up factor of \num{20} and resonance frequencies between \SI{68}{\mega\hertz} and \SI{72}{\mega\hertz}, which were driven by an amplified rf generator. The SNSPDs were current-biased using a low-noise voltage source and a \SI{10}{\kilo\ohm} series resistor through the low-frequency port of a bias tee. Output pulses from the SNSPD, emerging from the high-frequency port of the bias tee, were amplified, lowpass-filtered and notch filtered to remove any rf pickup before being sent to an oscilloscope or counter. The measured pulse decay time constant for the SNSPDs was \SI{\sim25}{\nano\second}, giving a device kinetic inductance of \SI{1.3}{\micro\henry}.

The measured bright count rates (BCR) and dark count rates (DCR) as a function of dc bias current $I_b$ are presented in Fig.~\ref{measurement} for different operating temperatures and mirror configurations, each at several values of applied rf voltage on the trap rf electrodes. Based on simulations and measurements in Ref.~\onlinecite{Todaro.2021}, rf at \SI{70}{\mega\hertz} with \SI{40}{\volt} amplitude on the trap electrodes should confine a single $^{40}$Ca$^+$ ion with axial (radial) secular frequencies of \SI{2}{\mega\hertz} ($\sim$\SI{5}{\mega\hertz}), with somewhat higher confinement directly over the SNSPD. Lower rf amplitudes should also be usable, and reducing the rf frequency will reduce the required rf amplitude proportionally to achieve the same confinement. 

The incident photon flux on the SNSPD is the same for all traces within a given panel, but not necessarily the same for different panels. The BCR is proportional to the system detection efficiency (SDE), the probability that a photon incident on the detector is registered as a count by the room-temperature electronics. The data with no applied rf potential show typical SNSPD behavior for all devices: a sigmoid-shaped SDE curve that rises and eventually plateaus with increasing bias current. The maximum dc bias current that can be applied before the SNSPD latches in the normal state, called the switching current, is denoted with a colored tick mark on the horizontal axis. The measured DCR is generally below 1 count per second until the bias current is within a few $\mu$A of the switching current, at which point the DCR is observed to increase exponentially with bias current. For ion trapping applications, the dark count rate due to stray light from the readout laser beam(s) is generally well above 1 count per second, so we did not attempt to characterize DCR below 1 count per second. This qualitative dark count behavior holds both with and without applied rf potential. 

Fig.~\ref{measurement}(a) plots BCR and DCR from an ion trap device with a large mirror under the SNSPD that is left electrically floating (not grounded). With increasing applied rf amplitude, the switching current is reduced and the shape of the BCR vs. $I_b$ curve flattens. These effects have been observed before in ion-trap-integrated SNSPDs, and are believed to arise from induced rf currents flowing in the SNSPD~\cite{Slichter.2017, Todaro.2021}. The switching current in the absence of rf potential is considerably larger than that of the devices in Refs.~\onlinecite{Slichter.2017, Todaro.2021}, so the SNSPD continues to operate even at much higher rf amplitudes than were previously demonstrated. However, the maximum achievable BCR (and thus maximum SDE) starts to degrade with increasing rf amplitude. By contrast, the devices in panels (b), (c), and (d) of Fig.~\ref{measurement}, where the mirror under the SNSPD is electrically grounded, show qualitatively different behavior. First, the reduction in switching current is generally smaller for a given rf amplitude than for the device with the ungrounded mirror in Fig.~\ref{measurement}(a). Second, the shape of the BCR vs. $I_b$ curve does not flatten, but instead shifts so that the BCR begins to rise at smaller bias currents. The maximum achievable BCR also remains unaffected for considerably higher rf amplitudes. 

\begin{figure}%[t]
    \centering
    \begin{subfigure}[b]{0.48\textwidth}
        \centering
        \includegraphics[width=1.0\textwidth]{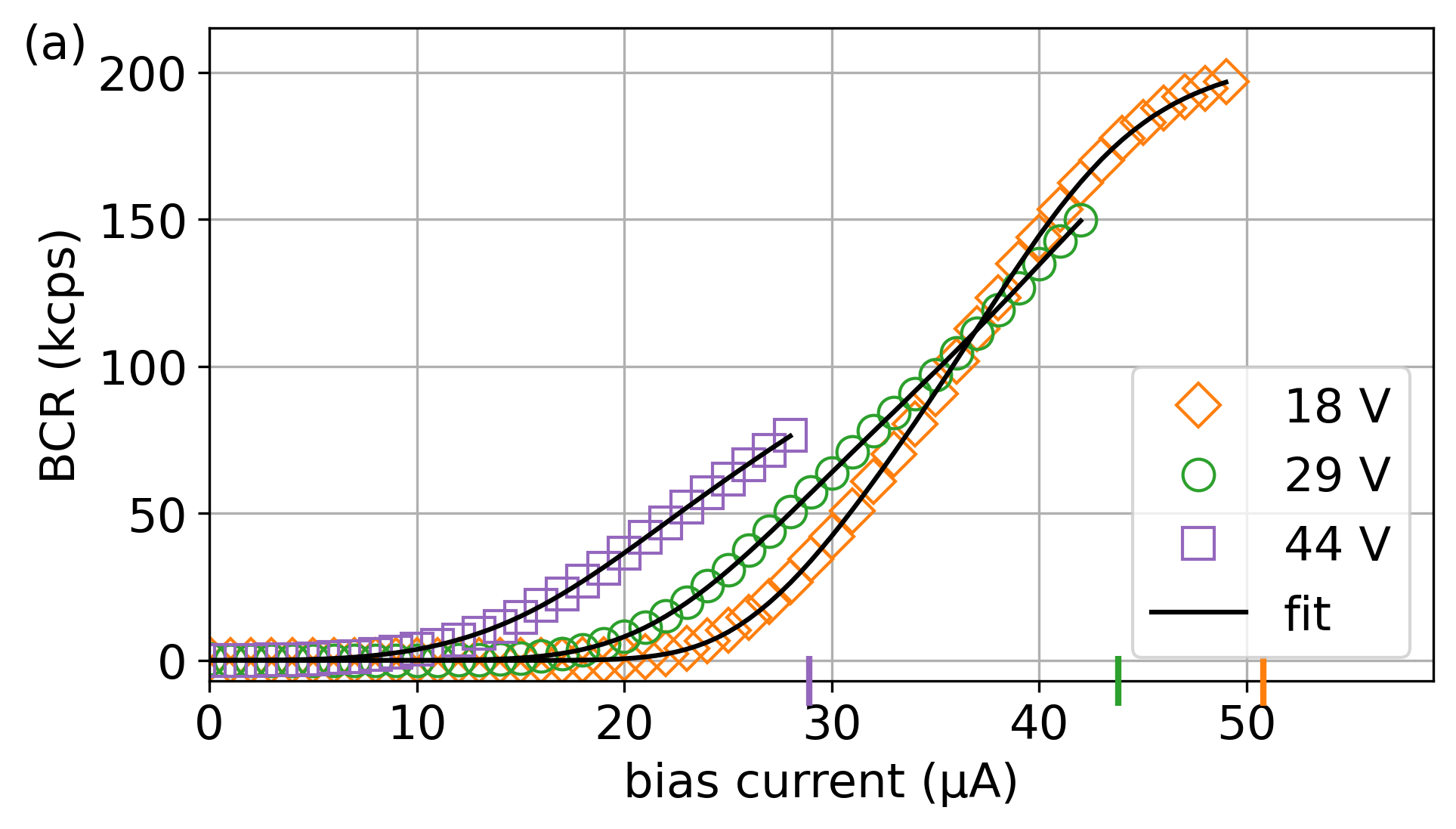}
        \vspace{-5mm}
    \end{subfigure}  
    \begin{subfigure}[b]{0.48\textwidth}
        \centering
        \includegraphics[width=1.0\textwidth]{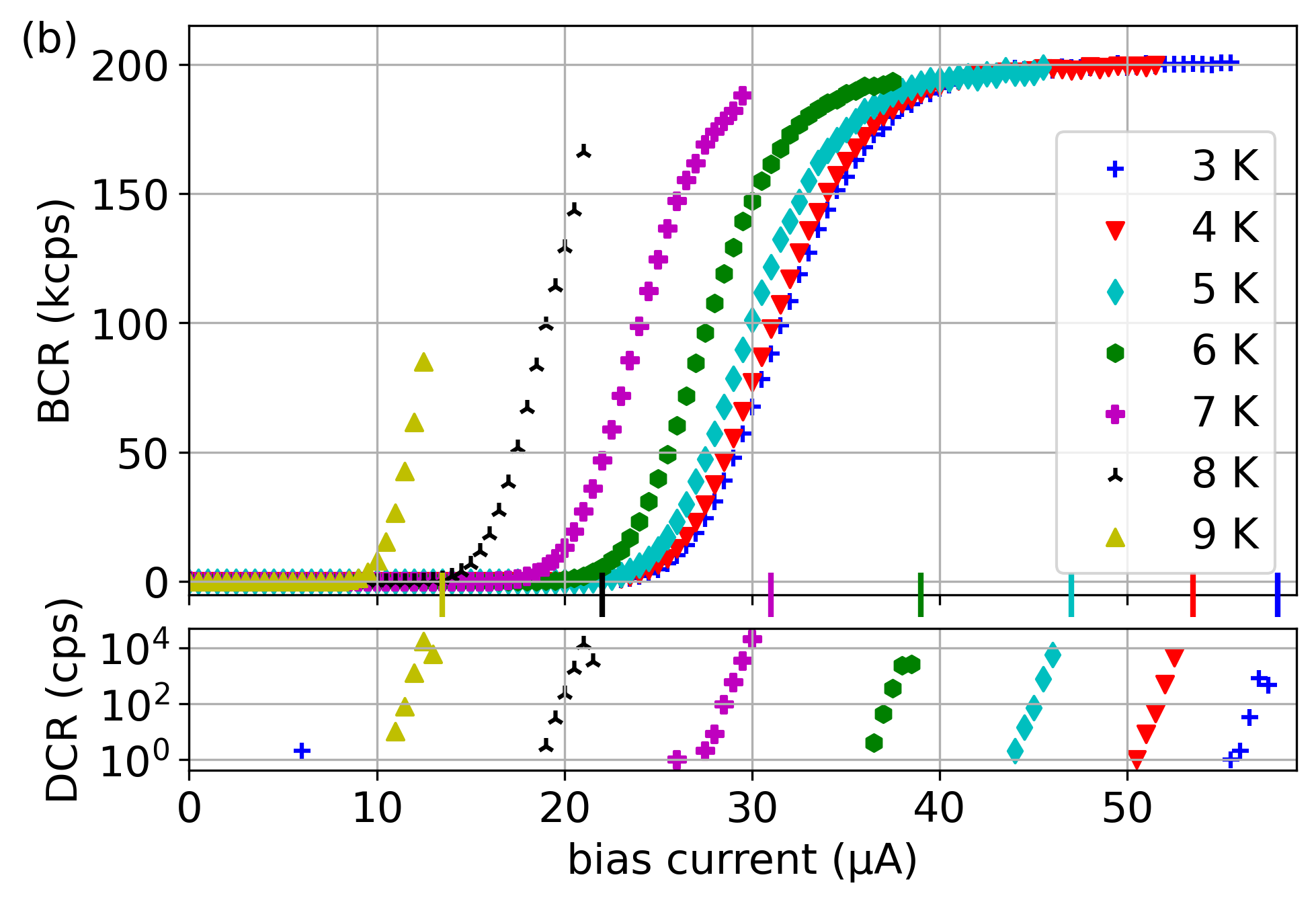}
        \vspace{-5mm}
    \end{subfigure}  
    \caption{Bright count rate (BCR) and dark count rate (DCR) in counts per second (cps) as a function of bias current. (a) Curve fits for an induced rf current model to the data in Fig.~\ref{measurement}(a). (b) BCR and DCR vs. $I_b$ without rf drive at different temperatures for a trap-integrated SNSPD. The switching currents are indicated with markers on the horizontal axis.}
    \label{temperature}
\end{figure}

In Fig.~\ref{temperature}(a), we fit the data from Fig.~\ref{measurement}(a) to a model for BCR vs. $I_b$ in the presence of induced rf currents in the SNSPD. This model accounts for both spatially uniform and spatially varying induced currents, and was presented in detail in Ref.~\onlinecite{Todaro.2021}. We expanded the model slightly to allow for small lateral shifts of the BCR vs. $I_b$ curve of the type seen in the other panels of Fig.~\ref{measurement}. We believe these shifts are due to local heating of the SNSPD from rf dissipation as described below. The strong agreement between data and fits suggests that induced rf currents in the SNSPD are present in the device with the non-grounded mirror. By contrast, this model does not fit well to any of the data from the devices with grounded mirrors, as the model indicates that induced rf currents always flatten the slope of the BCR vs. $I_b$ curve. However, the response of the SNSPDs to increasing rf is similar to their response to increasing temperature, shown in Fig.~\ref{temperature}(b) for a representative device. We therefore posit that on-chip power dissipation from the rf drive causes local heating of the SNSPD, and that this is the primary contribution to the observed rf dependence of the BCR versus bias current. We used an trap-integrated SNSPD in a different device to probe for local heating, using the SNSPD switching current as a proxy for its temperature. If the applied rf ($V_\mathrm{peak}=45$ V) is turned off and the switching current is measured in the ensuing \SI{\sim10}{\micro\second}, we infer an SNSPD temperature increase of roughly \SI{0.5}{\kelvin} over the steady-state temperature with no rf drive. This provides qualitative, but not quantitative, support for the local heating hypothesis. The rf electrodes in these devices are \SI{350}{\nano\meter} thick, as opposed to the \SI{\sim6}{\micro\meter} thickness in devices that would be used for ion trapping~\cite{Todaro.2021}; the corresponding increase in trace resistance and thus power dissipation may mean that local heating is more pronounced in these devices than in ones fabricated with thicker Au electrodes for use with ions, which may therefore exhibit lower sensitivity to applied rf potential.

%\begin{figure}%[t]
%    \includegraphics[width=0.5\textwidth]{figures/N2110291_CH2_2p6K_395nm_5uA_RF0dBm_2Filt_35uA_comb_paper.png}
%    \caption{Filtered and unfiltered SNSPD pulse shape of an ion trap chip with small, grounded mirror recorded with an oscilloscope at the output after amplification at T~=~\SI{2.6}{\kelvin}. The pulses are fitted to calculate the kinetic inductance.}
%    \label{pulse}
%\end{figure}

To determine a calibrated value of the detector SDE, we fabricated and tested stand-alone SNSPDs using the self-aligned design described in Ref.~\onlinecite{miller.2011}, where an optical fiber is butt-coupled to the SNSPD at cryogenic temperature. The layer stack and nanowire geometry was the same as for the trap-integrated devices, except the thickness of the bottom SiO\textsubscript{2} layer was optimized to a thickness of \SI{186}{\nano\meter} for theoretically maximum SDE. We illuminated the standalone SNSPDs with a fiber-coupled LED source with a peak wavelength of \SI{395}{\nano\meter}. The photon flux at the fiber output was calibrated at room temperature using a photomultiplier tube (PMT) whose quantum efficiency was reported by the manufacturer. The SDE of the SNSPDs could then be determined relative to that of the PMT by measuring the ratios of their respective count rates at a fixed LED power. The self-aligned SNSPD stack with an aluminum mirror reached a maximum SDE of \SI{68}{\percent}, while SNSPDs without mirror achieved an SDE of \SI{40}{\percent}, both measured at \SI{2.5}{\kelvin}. The SDE of SNSPDs well below the superconducting transition temperature is typically independent of temperature~\cite{wollman.2017}.

In summary, we have presented an improved, shielded NbTiN SNSPD integrated into a linear ion trap. These devices exhibit an order of magnitude increase in tolerance to applied trapping rf potential, along with an increase of at least \SI{2.5}{\kelvin} in operating temperature, relative to previous work with an identical trap electrode geometry~\cite{Todaro.2021}. These trap-integrated SNSPDs should be usable to perform fluorescence-based state readout of ion species with charge-to-mass ratios as low as $\sim$\num{0.01} elementary charges per unified atomic mass unit, encompassing most of the ion species typically used in quantum information experiments. Future work will target further improvements in SDE of the SNSPDs and further increases in trap rf tolerance through additional shielding and reduced on-chip rf dissipation, with the goal of using the trap-integrated SNSPDs for high-fidelity readout of qubits encoded in Ca$^+$ ions.

\vspace{5mm}

We thank D. V. Reddy and N. K. Lysne for thoughtful comments on the manuscript. This research was funded by NIST (https://ror.org/05xpvk416) and the IARPA LogiQ Program. This is a contribution of NIST, an agency of the U.S. Government, and is not subject to U.S. copyright.

\section*{Author Declarations}
\vspace{-5mm}
\subsection*{Conflict of Interest} 
\vspace{-5mm}
The authors have no conflicts to disclose.

\subsection*{Author Contributions} 
\vspace{-5mm}
B.H. performed the simulations, fabricated the devices, took the data, and analyzed the data with assistance from D.H.S. and V.B.V.; B.H. and D.H.S. wrote the manuscript with input from all authors; V.B.V. and D.H.S. supervised the work with assistance from D.L., R.P.M., and S.W.N.; V.B.V., D.H.S., D.L., R.P.M., and S.W.N. secured funding for the work.

\textbf{Benedikt Hampel:} Conceptualization (equal); Data curation (lead); Formal analysis (lead); Investigation (lead); Methodology (equal); Software (lead); Validation (equal); Visualization (lead); Writing - original draft (equal); Writing - review \& editing (equal).
\textbf{Daniel H. Slichter:} Conceptualization (equal); Formal analysis (supporting); Funding acquisition (equal); Investigation (supporting); Methodology (equal); Project administration (equal); Resources (equal); Software (supporting); Supervision (lead); Validation (equal); Writing - original draft (equal); Writing - review \& editing (equal). 
\textbf{Varun B. Verma:} Conceptualization (equal); Formal analysis (supporting); Funding acquisition (equal); Investigation (supporting); Methodology (equal); Project administration (equal); Resources (equal); Software (supporting); Supervision (lead); Validation (equal); Writing - review \& editing (equal).
\textbf{Dietrich Leibfried:} Funding acquisition (equal); Project administration (equal); Resources (equal); Supervision (supporting);   Writing - review \& editing (equal).
\textbf{Richard P. Mirin:} Funding acquisition (equal); Project administration (equal); Resources (equal); Supervision (supporting);   Writing - review \& editing (equal).
\textbf{Sae Woo Nam:} Funding acquisition (equal); Project administration (equal); Resources (equal); Supervision (supporting);  Writing - review \& editing (equal).

\section*{Data Availability}
\vspace{-5mm}
The data that support the findings of this study are openly available in the NIST Enterprise Data Inventory (EDI) at http://doi.org/[doi].

\section*{References}
% Create the reference section using BibTeX:
\bibliography{APL_IonTrap}

\end{document}